\documentclass[aps,prb,floatfix,showpacs,twocolumn,superscriptaddress]{revtex4}

\usepackage{color}
\usepackage{bm}
\usepackage{amssymb}
\usepackage{amsmath}
\usepackage{amstext}
\usepackage{graphics}
\usepackage{epsfig}
\usepackage{placeins}
\usepackage{psfrag}
\usepackage{subfigure}
\usepackage{times}

\newcommand{\bv}[1]
{
\left(
\begin{matrix}
#1
\end{matrix}
\right)
}

\begin{document}

\preprint{ Version \today}

\title{Harmonic oscillator model for current- and field-driven magnetic vortices}

\author{Benjamin Kr\"uger}

\affiliation{I. Institut f\"ur Theoretische Physik, Universit\"at Hamburg,
  Jungiusstr. 9, 20355 Hamburg, Germany}

\author{Andr\'e Drews}
\author{Markus Bolte}
\author{Ulrich Merkt}

\affiliation{Institut f\"ur Angewandte Physik und Zentrum f\"ur Mikrostrukturforschung, Universit\"at Hamburg,
  Jungiusstr. 11, 20355 Hamburg, Germany}

\author{Daniela Pfannkuche}

\affiliation{I. Institut f\"ur Theoretische Physik, Universit\"at Hamburg,
  Jungiusstr. 9, 20355 Hamburg, Germany}

\author{Guido Meier}

\affiliation{Institut f\"ur Angewandte Physik und Zentrum f\"ur Mikrostrukturforschung, Universit\"at Hamburg,
  Jungiusstr. 11, 20355 Hamburg, Germany}

\date{\today}

\begin{abstract}
In experiments the distinction between spin-torque and Oersted-field driven magnetization dynamics is still an open problem. Here, the gyroscopic motion of current- and field-driven magnetic vortices in small thin-film elements is investigated by analytical calculations and by numerical simulations. It is found that for small harmonic excitations the vortex core performs an elliptical rotation around its equilibrium position. The global phase of the rotation and the ratio between the semi-axes are determined by the frequency and the amplitude of the Oersted field and the spin torque.
\end{abstract}

\date{\today}


\maketitle

Recently it has been found that a spin-polarized current flowing through a magnetic sample interacts with the magnetization and exerts a torque on the local magnetization.~\cite{Berger1996,Slonczewski1996} A promising system for the investigation of the spin-torque effect is a vortex in a micro- or nanostructured magnetic thin-film element. Vortices are formed when the in-plane magnetization curls around a center region. In this few nanometer large center region~\cite{Wachowiak02}, called the vortex core, the magnetization turns out-of-plane to minimize the exchange energy.~\cite{ShinjoScience2000} It is known that these vortices precess around their equilibrium position when excited by magnetic field pulses~\cite{ChoeScience2004, Waeyenberge06} and it was predicted that spin-polarized electric currents can do the same.~\cite{ShibataPRB2006} The spacial restriction of the vortex core as well as its periodic motion around its ground state yield an especially accessible system for space- and time-resolved measurements with scanning probe and time-integrative techniques such as soft X-ray microscopy or X-ray photoemission electron microscopy.~\cite{Stoll04,Waeyenberge06,Guslienko06,ChoeScience2004,RaabePRL2005} Magnetic vortices also occur in vortex domain walls. The motion of such walls has recently been investigated intensively.~\cite{Meier07,KlaeuiPRL2005} Understanding the dynamics of confined vortices can give deeper insight in the mechanism of vortex-wall motion.~\cite{HePRB2006} An in-plane Oersted field accompanying the current flow also influences the motion of the vortex core. For the interpretation of experimental data it is crucial to distinguish between the influence of the spin torque and of the Oersted field.~\cite{Bolte07}

\begin{figure}
\includegraphics[angle = 0, width = 1.0\columnwidth]{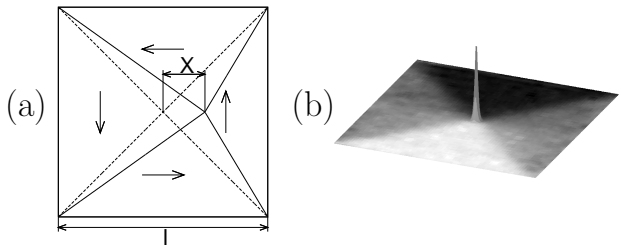}
\caption{(a) Scheme of the magnetization in a square magnetic thin-film element with a vortex that is deflected to the right. (b) Magnetization of a vortex in its static ground state. The height denotes the $z$-component while the gray scale corresponds to the direction of the in-plane magnetization. \label{model}}
\end{figure}

In this paper we investigate the current- and field-driven gyroscopic motion of magnetic vortices in square thin-film elements of size $l$ and thickness $t$ as shown in Fig.~\ref{model} and present a method to distinguish between spin torque and Oersted field driven magnetization dynamics. In the presence of a spin-polarized current the time evolution of the magnetization is given by the extended Landau-Lifshitz-Gilbert equation
\begin{equation}
\begin{split}
\frac{d \vec M}{d t} =
 &- \gamma \vec M \times \vec  H_{\mbox{eff}} + \frac{\alpha}{M_s} \vec M \times \frac{d \vec M}{d t}\\
 &- \frac{b_j}{M_s^2} \vec M \times \left( \vec M \times (\vec j \cdot \vec \nabla) \vec M \right)\\
 &- \xi \frac{b_j}{M_s} \vec M \times (\vec j \cdot \vec \nabla) \vec M
\end{split}
\label{LLG-Zang-Li}
\end{equation}
with the coupling constant $b_j = P \mu_B/[e M_s (1+ \xi^2)]$ between the current and the magnetization where $P$ is the spin polarization, $M_S$ the saturation magnetization, and $\xi$ the degree of non-adiabaticity.~\cite{Zhang04} If the vortex keeps its static structure, its motion with the velocity $\vec v$ can be described using the Thiele equation.~\cite{Thiele1974} This equation was expanded by Nakatani~et~al.~\cite{ThiavilleNatureMat} to include the action of a spin-polarized current flowing in the sample,
\begin{equation}
\vec F + \vec G \times ( \vec v + b_j \vec j) + D ( \alpha \vec v + \xi b_j \vec j) = 0.
\label{eq_Thile}
\end{equation}
Denoting the out-of-plane angle of the magnetization with $\theta$ and the angle of the in-plane magnetization with $\phi$, the force due to the external and the stray field is
\begin{equation}
\vec F = - \mu_0 \int dV \, \left[ (\vec \nabla \theta) \frac{\partial}{\partial \theta}+ (\vec \nabla \phi) \frac{\partial}{\partial \phi} \right] (\vec H_{sz} \cdot \vec M).
\end{equation}
The gyrovector
\begin{equation}
\begin{split}
\vec G & = - \frac{M_s \mu_0}{\gamma} \int dV \, \sin(\theta) (\vec \nabla \theta \times \vec \nabla \phi) \\ & = - \frac{2 \pi M_s \mu_0 t p}{\gamma} \vec e_z = G_0 \vec e_z,
\end{split}
\end{equation}
indicates the axis of precession and points out-of-plane. The dissipation tensor is given by
\begin{equation}
D = - \frac{M_s \mu_0}{\gamma} \int dV \, (\vec \nabla \theta \vec \nabla \theta + \sin^2(\theta) \vec \nabla \phi \vec \nabla \phi).
\end{equation}
It is diagonal with
\begin{equation}
D_{xx} = D_{yy} = D_0 \approx - \frac{\pi M_s \mu_0 t \ln(l/a)}{\gamma}, \; \; \; \; \; \; D_{zz} = 0.
\end{equation}
The constant $a$ is the lower bound of the integration. It is in the order of magnitude of the radius of the vortex core.~\cite{Wachowiak02,Huber82,HePRB2006,Guslienko05} A polarization $p$ of $+1$ ($-1$) denotes that the magnetization in the vortex core is parallel (antiparallel) to the $z$-axis. The velocity of the vortex core is in-plane and hence perpendicular to the gyrovector. Thus Eq.~(\ref{eq_Thile}) can be rewritten as
\begin{equation}
\vec G \times \vec F - G_0^2 ( \vec v + b_j \vec j) + D_0 \vec G \times ( \alpha \vec v + \xi b_j \vec j) = 0.
\label{eq_Thile_cross}
\end{equation}
By calculating $\vec G \times \vec v$ from Eq.~(\ref{eq_Thile_cross}) and inserting the result in Eq.~(\ref{eq_Thile}) we can derive the velocity
\begin{equation}
\begin{split}
(G_0^2 + D_0^2 \alpha^2) \vec v & = \vec G \times \vec F - D_0 \alpha \vec F - (G_0^2 + D_0^2 \alpha \xi) b_j \vec j\\
 & + b_j D_0 \vec G \times \vec j (\xi - \alpha)
\end{split}
\label{eq_velocity}
\end{equation}
of the vortex core. As for any square-symmetric confining potential, the stray-field energy for small deflections can be modeled as a parabolic potential
\begin{equation}
E_s = \frac{1}{2} m \omega_r^2 (X^2 + Y^2)
\end{equation}
with the coordinates $X$ and $Y$ of the vortex core (see Fig.~\ref{model}a).

In the following a spacially homogeneous current in $x$-direction is investigated. Due to possible inhomogeneities in real samples the current flow may vary in the out-of-plane direction. This results in an in-plane Oersted field which is perpendicular to the direction of the current flow. In the following this Oersted field is accounted for by a homogeneous magnetic field in $y$-direction. Both driving forces may depend on time. To estimate the Zeeman energy due to the Oersted field $H$, the magnetization pattern is divided into four triangles (see Fig.~\ref{model}a). Assuming that the magnetization is uniform in each of these triangles the total Zeeman energy is given by
\begin{equation}
E_z = \frac{\mu_0 M_s H l t c}{2} \left[ \left( \frac{l}{2}+X \right) - \left( \frac{l}{2}-X \right) \right],
\label{eq_Zeeman_energy}
\end{equation}
with the chirality $c$ of the vortex. A chirality of $+1$ ($-1$) denotes a counterclockwise (clockwise) curling of the magnetization around the vortex core. We will see that this simple approximation describes the field-induced vortex motion sufficiently well. In this case the force is given by
\begin{equation}
\vec F = - \vec \nabla (E_s + E_z) = -\mu_0 M_s H l t c \vec e_x - m \omega_r^2 X \vec e_x - m \omega_r^2 Y \vec e_y.
\label{eq_force}
\end{equation}
Inserting Eq.~(\ref{eq_force}) in Eq.~(\ref{eq_velocity}) yields the equation of motion for the vortex. In the absence of current and field the excited vortex performs an exponentially damped spiral rotation around its equilibrium position with its free frequency
\begin{equation}
\omega = - \frac{p G_0 m \omega_r^2}{G_0^2 + D_0^2 \alpha^2}
\label{eq_omega_free}
\end{equation}
and the damping constant
\begin{equation}
\Gamma = - \frac{D_0 \alpha m \omega_r^2}{G_0^2 + D_0^2 \alpha^2}.
\label{eq_gamma}
\end{equation}
From Eqs. (\ref{eq_omega_free}) and (\ref{eq_gamma}) one easily obtains that
\begin{equation}
D_0 \alpha = \frac{\Gamma p G_0}{\omega}.
\label{eq_omega_free_gamma}
\end{equation}
For thin-film systems ($t/l \lesssim 0.1$) the resonance frequency of a vortex is proportional to the inverse lateral dimension $1/l$.~\cite{Guslienko02} Here, we obtain from Eq.~(\ref{eq_omega_free_gamma}) that the damping constant $\Gamma$ also has a characteristic length dependence, $\Gamma \propto \ln(l/a)/l$. Substituting $D_0 \alpha$ using Eq.~(\ref{eq_omega_free_gamma}) the equation of motion of the vortex can be written as
\begin{equation}
\begin{split}
\bv{\dot X\\ \dot Y} & = \bv{- \Gamma & -p \omega\\ p \omega & -\Gamma} \bv{X\\ Y}\\  & + \bv{\frac{p \omega \Gamma}{\omega^2 + \Gamma^2} \frac{\mu_0 M_s H l t c}{G_0} - b_j j - \frac{\Gamma^2}{\omega^2 + \Gamma^2} \frac{\xi - \alpha}{\alpha} b_j j\\ - \frac{\omega^2}{\omega^2 + \Gamma^2} \frac{\mu_0 M_s H l t c}{G_0} + \frac{p \omega \Gamma}{\omega^2 + \Gamma^2} \frac{\xi - \alpha}{\alpha} b_j j}.
\end{split}
\label{eq_diff_velocity_tmp}
\end{equation}
In the following we assume harmonic excitations, i.e., the magnetic field and the electrical current are of the form $H(t) = H_0 e^{i \Omega t}$ and $j(t) = j_0 e^{i \Omega t}$. The magnetic (Oersted) field and the electrical current are in phase. Assuming that the squared Gilbert damping is small ($\alpha^2 \ll 1$), the damping constant of the vortex is small compared to its frequency ($\Gamma^2 \ll \omega^2$). Then Eq.~(\ref{eq_diff_velocity_tmp}) has the solution
\begin{equation}
\begin{split}
\bv{X\\ Y} & = A \bv{i \\ p} e^{- \Gamma t + i \omega t} + B \bv{-i \\ p} e^{- \Gamma t - i \omega t} \\ & -\frac{e^{i \Omega t}}{\omega^2 + (i \Omega + \Gamma)^2} \\  & \times \bv{ \left( \tilde H + \frac{\Gamma}{\omega} \frac{\xi}{\alpha} \tilde j \right) \omega & + & \left( \frac{\Gamma}{\omega} \tilde H + \tilde j \right) i \Omega
\\ \tilde j \omega p & - & \left( \tilde H + \frac{\Gamma}{\omega} \frac{\xi - \alpha}{\alpha} \tilde j \right) i \Omega p},
\end{split}
\label{eq_solution}
\end{equation}
with $\tilde H = \gamma H_0 l c/(2 \pi)$ and $\tilde j = b_j j_0$. The first two terms with prefactors A and B are exponentially damped and depend on the starting configuration. Independent of the source of excitation, i.e., field or current, the sense of rotation of the vortex is given by its polarization, i.e., $p=+1$ ($p=-1$) denotes a counterclockwise (clockwise) rotation of the vortex core. Changing the sign of the chirality has the same effect as turning the magnetic field by $180^{\circ}$. Similar to the motion of magnetic domain walls in thin nanowires~\cite{Krueger07} the vortex is driven by the current and the magnetic field as well as by their time derivatives.

\begin{figure}
\includegraphics[angle = 270, width = 1.0\columnwidth]{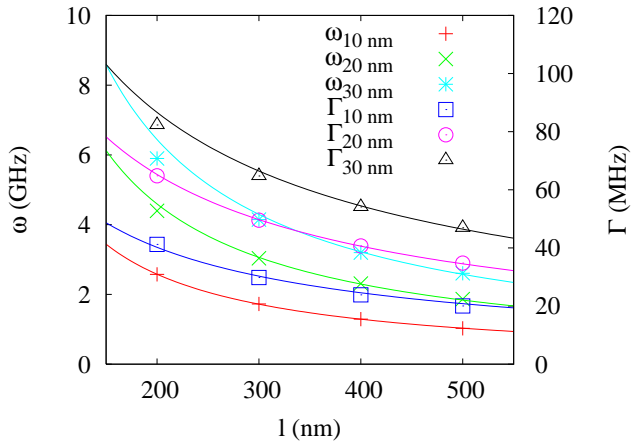}
\caption{(Color Online) Dependence of the frequency $\omega$ and the damping constant $\Gamma$ on the length $l$ for various thicknesses $t$ of the system. The symbols denote numerical results while the lines are fits with the analytical results. \label{frequency_fit}}
\end{figure}
\begin{figure}
\includegraphics[angle = 270, width = 1.0\columnwidth]{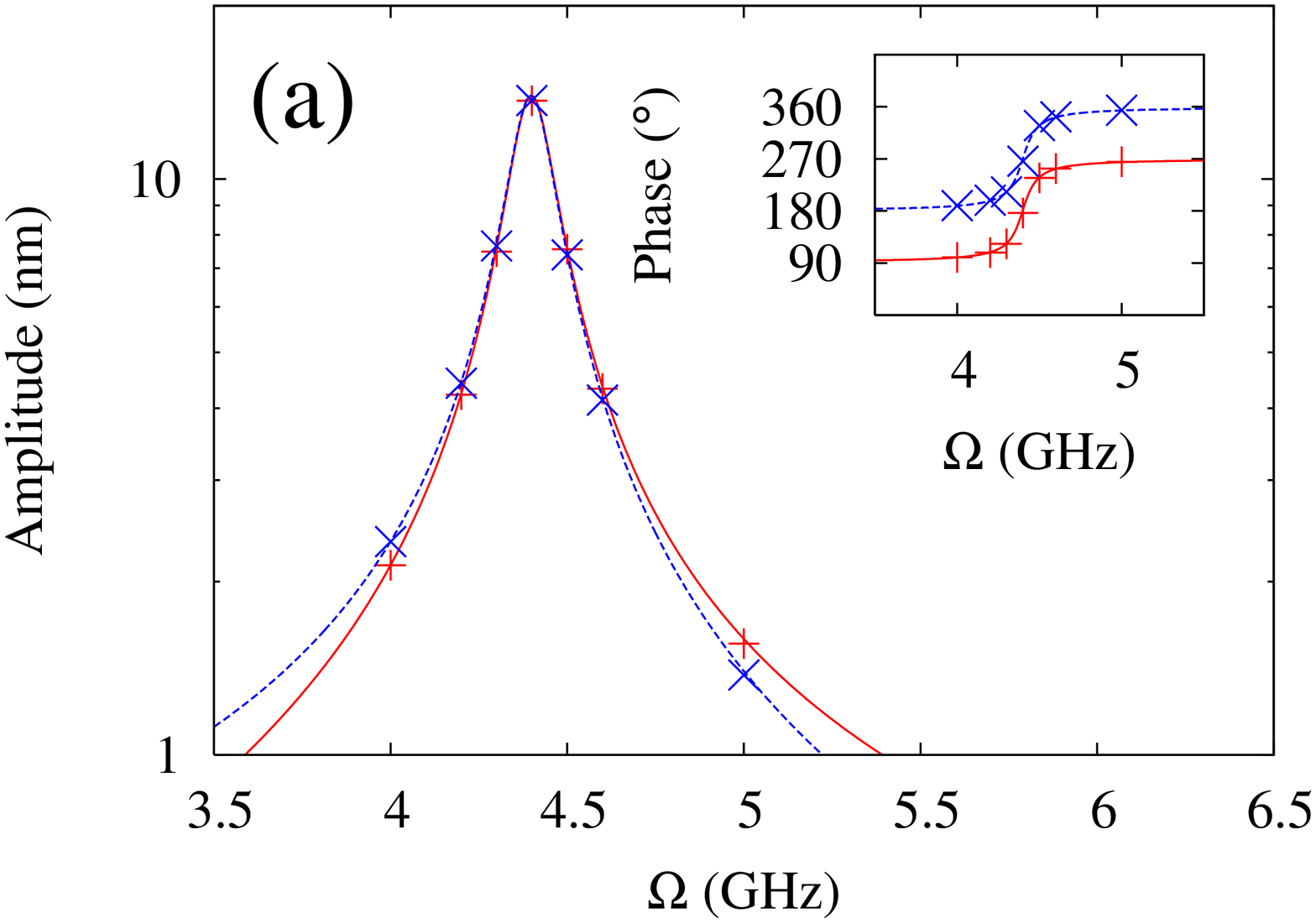}
\includegraphics[angle = 270, width = 1.0\columnwidth]{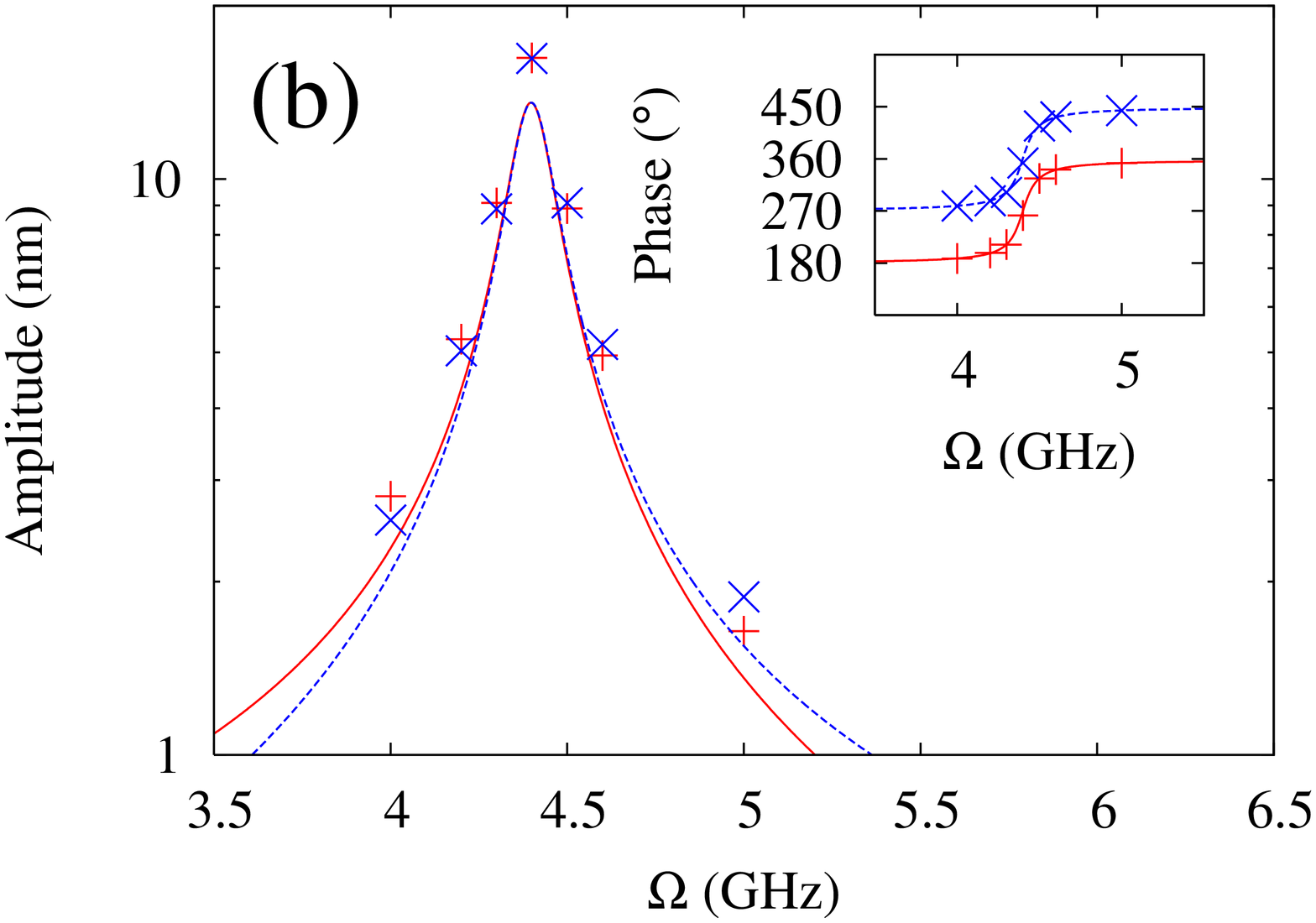}
\caption{(Color Online) Amplitude of the (a) current-driven and (b) field-driven vortex oscillation in $x$-direction (solid red line, pluses) and $y$-direction (dashed blue line, crosses) for a spin-polarized current density of $j P = 2.5 \cdot 10^{10}$~A/m$^2$ and a field of $H = 250$~A/m.  The insets show the phases between the maximum of the applied current or field and the core displacement in $x$-direction (solid red line, pluses) and $y$-direction (dashed blue line, crosses). The symbols denote numerical results while the lines are derived from the analytical expression in Eq.~(\ref{eq_solution}). \label{amplitude}}
\end{figure}

At resonance the amplitude of the vortex core displacement in $x$- and $y$-direction is the same and the vortex performs a circular rotation. A vortex which is excited with a non-resonant frequency has an elliptic trajectory. The ratio between the semi-axes is given by the ratio between the frequency of the excitation and the resonance frequency.

To test the applicability of the approximations leading to the analytical result in Eq.~(\ref{eq_solution}) we performed micromagnetic simulations for magnetic thin-film elements with different lengths, thicknesses, polarizations, and chiralities. The material parameters of permalloy are used, i.e., an exchange constant of $A = 13 \cdot 10^{-12}$ J/m and a saturation magnetization of $M_s = 8 \cdot 10^5$ A/m. For the Gilbert damping we use a value of $\alpha = 0.01$ which is in the regime as found by recent experiments.~\cite{Nibarger03,Schneider05,Liu07} The degree of non-adiabaticity $\xi$ is chosen to be equal to $\alpha$.~\cite{Meier07, Hayashi2006}

For the micromagnetic simulations we extended the implementation of the Landau-Lifshitz-Gilbert equation in the Object Oriented Micro Magnetic Framework (OOMMF) by the additional current-dependent terms of Eq.~(\ref{LLG-Zang-Li}).~\cite{OOMMF, Krueger07} The simulation cells are 2~nm in $x$- and $y$-direction which is well below the exchange length of permalloy. One cell of thickness $t$ was used in $z$-direction. As in the analytical model we substitute the Oersted field by a homogeneous magnetic field.

At first the four ground states with $c \pm 1$ and $p \pm 1$ are calculated for each $l$ and $t$. The ground states are then excited by a short current pulse. The frequency $\omega$ and the damping constant $\Gamma$ are obtained by fitting the subsequent free oscillation with the first two terms in Eq.~(\ref{eq_solution}). Results are presented in Fig.~\ref{frequency_fit} and exhibit a good agreement between the analytical model and the micromagnetic simulations.~\cite{note_a}

\begin{figure}
\includegraphics[angle = 270, width = 1.0\columnwidth]{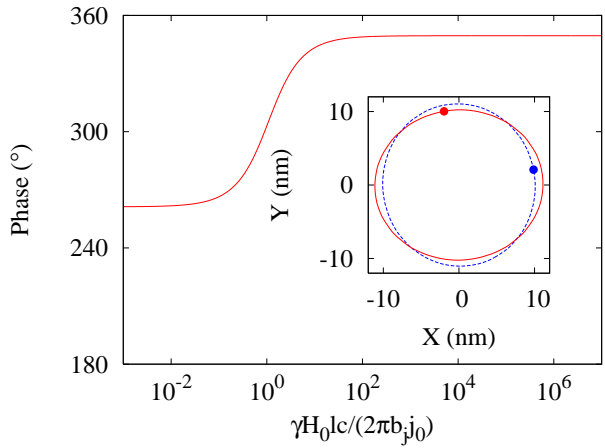}
\caption{(Color Online) Analytically calculated phase between the maximum current or magnetic field and the $x$-deflection of the vortex core for a 200~nm x 200~nm x 20~nm permalloy square excited with a frequency of $\Omega = 4.8$~GHz (above the resonance frequency of $\omega = 4.4$~GHz).
The inset shows a section of the sample with the simulated trajectories of the vortex core excited with i) (solid red line) a spin-polarized current density with an amplitude of $j P = 1.2 \cdot 10^{11}$~A/m$^2$  and ii) (dashed blue line) a magnetic field with an amplitude of $H = 1000$~A/m. Points denote the position of the vortex at maximum current (i) and magnetic field (ii), respectively.
\label{spacial}}
\end{figure}

For the driven oscillation we choose a magnetic film element with length $l = 200$~nm and thickness $t = 20$~nm. This system size allows for reasonable computing time. The magnetization is excited with harmonic currents with a spin-polarized current density $j P = 2.5 \cdot 10^{10}$~A/m$^2$ in $x$-direction. The field excitation was performed with a harmonic field of $H = 250$~A/m in $y$-direction. The amplitudes and the phases of the oscillation in $x$- and $y$-direction of a vortex with positive polarization and chirality are depicted in Fig.~\ref{amplitude}. In the current-driven oscillation an excellent accordance between analytical calculations and numerical simulations is found. In the field-driven case the amplitudes of the analytical solution are smaller than the amplitudes obtained from the micromagnetic simulations. These deviations are caused by the differences between the approximate magnetization depicted in Fig.~\ref{model} and the exact state. The phases between the maximum of the exciting magnetic field and the maximum deflection in $x$- and $y$-direction agree very well. Vortices with other polarization and chirality (not shown) yield the same accordance.~\cite{Yamada_note} 

From Eq.~(\ref{eq_solution}) one can see that the current and field induced forces on the vortex are of the same form. For experiments it is important to separate the Oersted-field and the spin-torque driven case. We describe the ratio between the field and current-induced forces on the vortex by $\tan \zeta = F_{\mbox{\scriptsize Oe}}/F_{\mbox{\scriptsize st}}$, i.e., a mixing angle of $\zeta = 0$ and $\zeta = \pm \pi/2$ denote the fully spin-torque driven and the fully field-driven case, respectively. There are two possibilities to determine the ratio of both forces. On the one hand for non-resonant excitations the trajectory of the vortex core is elliptical as illustrated in Fig.~\ref{spacial}. According to Eq. (\ref{eq_solution}) the direction of the major axis of the ellipse is determined by $\zeta$. The amplitude of the vortex motion decreases very fast when the excitation frequency deviates from resonance, i.e., for experimental observation very high current densities with frequencies close to resonance are needed. On the other hand the excitation mechanisms can be distinguished using the phase of the vortex deflection.~\cite{Bolte07} As indicated by the dots in Fig.~\ref{spacial} the position of the vortex at maximum current depends on $\zeta$, which can be determined from Eq.~(\ref{eq_solution}). The latter method is also applicable with excitations at resonance frequency.

In conclusion we derived an analytical expression for the current- and field-driven trajectory of a vortex in thin-film elements. The analytical result is compared to micromagnetic simulations. The accordance between both approaches is very good. The analytical expression enables us to determine the ratio between spin torque and Oersted field driven motion.

\begin{acknowledgments}
Financial support by the Deutsche Forschungsgemeinschaft via SFB 668 "Magnetismus vom Einzelatom zur Nanostruktur" and via Graduiertenkolleg 1286 "Functional metal-semiconductor hybrid systems" is gratefully acknowledged.
\end{acknowledgments}


\begin{thebibliography}{29}
\bibitem[{not()}]{note_a}
\bibinfo{note}{Note: $a$ is a fit parameter. The values used are 8.85~nm,
  11.66~nm, and 13.59~nm for film thicknesses of 10~nm, 20~nm, and 30~nm,
  respectively}.
\end{thebibliography}

\end{document}